\documentclass[aps,prb,groupedaddress,showpacs]{revtex4}
\usepackage{graphicx}

\begin{document}

\title{Andreev scattering and cotunneling between two 
superconductor-normal metal interfaces : the dirty limit}

\author{D. Feinberg}
\email[]{feinberg@grenoble.cnrs.fr}
\affiliation{Laboratoire d'Etudes des Propri\'et\'es Electroniques
des Solides, Centre National de la Recherche Scientifique$^{¤}$, BP 166, 38042 Grenoble Cedex
9, France}

\date{\today}

\begin{abstract}
Crossed Andreev reflections and cotunneling occur between two neighbouring 
superconductor-
normal metal or superconducting-ferromagnet interfaces. Previous works 
assumed a clean BCS superconductor. Here the calculation
of the corresponding crossed conductance terms is generalized to a dirty 
superconductor. The range of the effect is
shown to be the coherence length $\tilde{\xi} = \sqrt{\hbar D/\Delta}$, 
instead of the BCS coherence length $\xi_0$. Moreover, in three
dimensions, the algebraic prefactor scales 
as $1/r$ instead of $1/r^2$. The calculation involves the virtual 
diffusion probability of quasiparticles below the superconducting gap, in 
the normal
and the anomalous channel. 
\end{abstract}


\maketitle

\section{Introduction}
Hybrid structures involving superconductors and normal or ferromagnetic 
metals have received considerable 
interest in the context of spintronics \cite{review}. More recently, 
multiterminal structures were considered, where two metallic
leads are connected at a small distance to the same superconductor 
\cite{flatte,DF,jedema,falci,melin2,melin3,melin1,stm}. Coherent scattering may occur
between those leads, leading to original crossed conductance channels  \cite{lambert1,lambert2,falci}. 
The first one
generalizes Andreev reflection : an electron (resp. hole) incident on either 
contact is reflected as a hole (resp. electron) in the other one. This amounts to 
having a Cooper pair transferred to (from) the superconductor, each electron of the pair
passing at a different contact in the same direction. One may also have an electron (hole)
reflected as an electron (hole) from one contact to the other. This process which 
generalizes normal reflection has been named cotunneling since, in presence of the 
superconducting gap, a quasiparticle propagates 
in the superconductor as an evanescent state, in a way similar to what happens 
in presence of Coulomb blockade \cite{cotunneling}. Notice that here cotunneling is essentially 
elastic. The calculation of the 
scattering amplitudes and the corresponding
non-local conductances was performed on Refs. \cite{falci,melin1} in the 
case of a clean BCS superconductor. Normal reflections lead to
cotunneling, which conserves spin, while anomalous reflections lead to 
crossed Andreev conductance involving
opposite spin channels in the two leads. These effects give rise to a 
variety of new phenomena and potential applications. On one hand, 
when the leads are spin-polarized, the symmetry between these two 
processes is broken and interesting nonlocal 
magnetoconductance effets have been predicted \cite{DF,falci}. They can be 
used as a novel principle for a spin-sensitive STM \cite{stm}. 
On the other hand, crossed Andreev processes, as they lead to spatially 
separated singlet pairs, have signatures in crossed noise correlations 
\cite{torres,samuelsson,sanchez,pistolesi}. They can also be used
as a source of entangled electron pairs, a crucial resource for the 
treatment of quantum information \cite{loss,martin}. 

In order to maximize the crossed conductance effects, 
it is essential to optimize the physical regimes, the parameters and the geometry. 
For instance, for point contacts, the dependence of the crossed conductances
 with the contact distance $r$ is found to be 
$\sim (\frac{1}{k_Fr})^2 e^{-2r/\pi\xi_0}$ where $k_F$ is the Fermi 
wavevector and $\xi_0$ the BCS coherence length \cite{falci,loss}.
 This result is valid for a clean three-dimensional superconductor, and 
the algebraic prefactor reduces very strongly the amplitude of
the effect for realistic distances. For a $2D$ (resp. $1D$) 
superconductor, the exponent in the prefactor
is instead found to be $1$ (resp. $0$), offering a neat advantage. This 
lead to the proposal of inducing superconductivity in
 carbon nanotubes, in order to reach an effective $1D$ geometry 
 \cite{recher,bouchiat}. Another possibility is to use extended contacts, 
 which cancels the prefactor (see for instance \cite{stm}).

In the present work, the calculation of the crossed conductances is 
generalized to a diffusive superconductor. If the mean-free
path $l$ is smaller than the bare coherence length $\xi_0$, the question 
is : which one of the typical lengths, $l$ or the coherence length 
$\xi = \sqrt{\hbar D/\Delta} \sim \sqrt{\xi_0 l}$, controls the range of 
the effect ? The 
calculation involves the diffusion of (evanescent) quasiparticles between 
the two contacts. As a result, the range is found to be 
of order $\xi$. Although not really surprising, this result is not obvious a priori: 
$\xi$ is known to be the typical length for the variations of the superconducting
order parameter, while we are interested here into the damping length 
for quasiparticles. In other terms, it is not clear that in the diffusive regime
 the spatial 
dependence of the single particle propagator and of the local pair amplitude are 
governed by the same length.
Moreover, we find that in a three-dimensional dirty superconductor the 
prefactor becomes $(k_Fl)^{-1}(k_Fr)^{-1}$ 
instead of $(k_Fr)^{-2}$. Owing to the crucial importance of the prefactor, this 
noticeably increases the crossed effects as compared to a clean system.

This paper is organized as follows. In Section 1, using a tunneling 
Hamiltonian, the general result for the current across one of 
two neighbouring S/N interfaces is recalled. In Section 2, the calculation 
is performed for 
a dirty superconductor, within the lowest order approximation. The effect 
of dimensionnality and geometry are 
discussed at the end of the paper.

\section{Clean superconductor : tunneling interfaces}

\begin{figure}
\includegraphics[scale=.5]{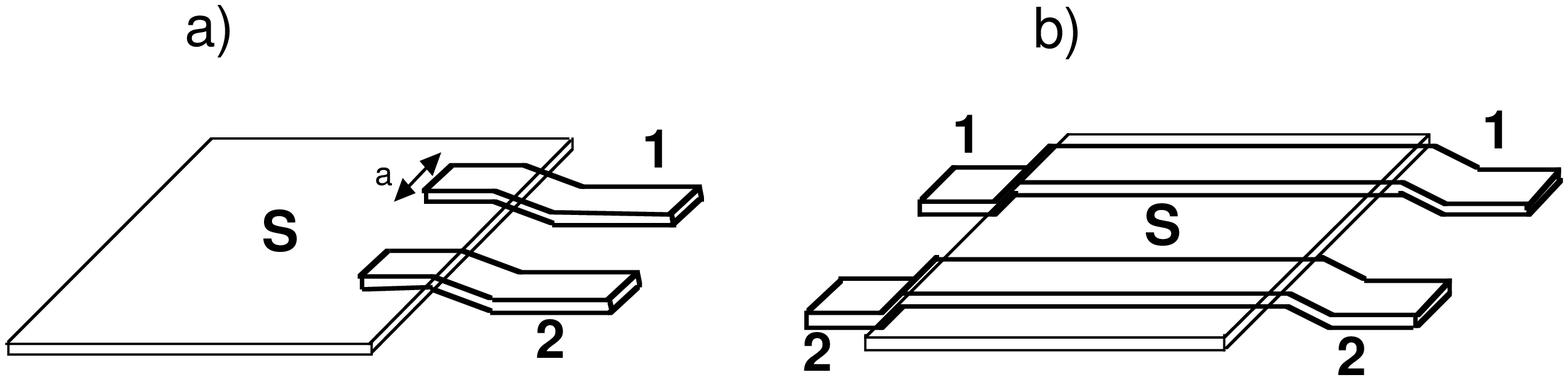}
\caption{Schema of a set-up with two neighboring S/N junctions. $1$ and $2$ 
denote normal leads; a) point-like contacts; b) extended contacts} 
\end{figure}

Let us consider a ballistic BCS superconductor, connected to two leads 1 
and 2 (depicted on Figure 1), with voltages $V_1$ and $V_2$ with respect 
to the superconductor, by tunneling contacts described by the Hamiltonian : 

\begin{eqnarray}
{\cal H}_{T1} \;=\;
\sum_{kp\sigma} \; T^1_{kp} \; c^{\dagger}_{k\sigma} d_{p\sigma} \,+\,
H. c.
\quad ; \quad
{\cal H}_{T2} \;=\;
\sum_{pq\sigma} \; T^{2}_{pq} \; d^{\dagger}_{p\sigma} c_{q\sigma} \,+\,
H. c.
\end{eqnarray}

\noindent
where $T^1_{kp}$ and $T^{2}_{qp}$ are matrix elements (hereafter assumed to be equal to 
$T_1$, $T_2$) 
between single electron states $k \in 1$,  $p \in S$ and $q \in 2$. 

Let us first consider single channel leads, at $T=0$. 
Here we limit ourselves
to lowest order results, which can also be obtained by the Keldysh technique 
\cite{melin2,melin3,melin1,cuevas} or the golden rule 
approximation \cite{falci,hekking}. Dropping the 
usual Andreev reflection current \cite{hekking}, we focus on the nonlocal 
contributions, e. g. the current induced in one lead by the voltage applied 
on the other. Using the fact that the spectral 
functions $g_i$'s ($i=1,2$) in the metallic leads decay on
the scale of the Fermi wavelength, one obtains \cite{falci} for the "Crossed Andreev" current 
$I_{CAnd}$ and the "Elastic Cotunneling" current $I_{ECot}$ 

\begin{eqnarray}
I_{CAnd} = \sum_{\sigma}\frac{4\pi^2 e}{h} |T^{1}T^{2}|^{2} \int \; d\omega 
\;\; \Xi_{12}^{And}(\omega, \sigma) \; [n_F(\omega-eV_1) - n_F(\omega + 
eV_2)]\\
I_{ECot} =\sum_{\sigma}\frac{4\pi^2 e}{h} |T^{1}T^{2}|^{2} \int \; d\omega 
\;\; \Xi_{12}^{Cot}(\omega, \sigma)\;[n_F(\omega-eV_1) - n_F(\omega - 
eV_2)]
\end{eqnarray}

\noindent
with

\begin{eqnarray}
\Xi_{12}^{And}(\omega, \sigma)\;=\;\int_1 \;d\vec{r}_1 \;\int_2 \;d\vec{r}_2\;\;
f^r_{\sigma}(\omega,r_{12})f^a_{\sigma}(\omega,r_{21}) \; g_{1 
\sigma}(\omega) g_{2 -\sigma}(-\omega)\\
\Xi_{12}^{Cot}(\omega, \sigma)\;=\;\int_1 \;d\vec{r}_1 \;\int_2 \;d\vec{r}_2\;\;
 g^r_{\sigma}(\omega,r_{12})g^a_{\sigma}(\omega,r_{21}) \; g_{1 
\sigma}(\omega) g_{2 \sigma}(\omega)
\end{eqnarray}

\noindent
where the integrals run on the contact areas, 
$r_{ij}=|\vec{r}_{i}-\vec{r}_{j}|$, $g^{r}_{\sigma}(\omega, r_{ij})$ 
  and $f^{r}_{\sigma}(\omega, r_{ij})$ 
are respectively the time Fourier transforms of the normal 
$-i\langle T\,\{c_{i\sigma}(t),c^{\dagger}_{j\sigma}(0)\}\rangle$ 
and anomalous $i\langle T\,\{c_{i\sigma}(t),c_{j\sigma}(0)\}\rangle$ 
bare retarded Green's functions in the superconductor. 
Those are given in three dimensions by

\begin{eqnarray}
g^r(r,\omega)\;=\;-\frac{m}{2\pi\hbar^2}\;\frac{1}{r}\;e^{-r/2\xi(\omega)}\;
[sink_Fr\;\frac{\omega}{\sqrt{\Delta^2-\omega^2}}\;+\;cosk_Fr]\\
f^r(r,\omega)\;=\;-\frac{m}{2\pi\hbar^2}\;\frac{1}{r}\;e^{-r/2\xi(\omega)}\;
sink_Fr\;\frac{\Delta}{\sqrt{\Delta^2-\omega^2}}
\end{eqnarray}

\noindent
where $\xi_{\omega} = \xi_0 \frac{\Delta}{\sqrt{\Delta^2-\omega^2}}$ is a 
generalized frequency-dependent coherence length. 
One can then calculate the conductances associated
respectively to crossed Andreev and elastic cotunneling processes, e.g. 
$G_{CAnd} = dI_{CAnd}/d(V_1 + V_2)$ and $G_{ECot} = 
dI_{ECot}/d(V_1 - V_2)$. Up to geometrical factors, 
the result for interfaces of radius $a << \xi_0$ but much larger than the Fermi length 
$k_F^{-1}$ and distant by $r>>a$ (Figure 1a)
is \cite{falci}

\begin{eqnarray}
G_{CAnd} \sim \frac{h}{8e^2} \sum_{\sigma} 
\,G_{1\sigma}G_{2-\sigma}\;\frac{e^{-2 r/ \pi \xi_{\omega}}}{(k_Fr)^2}\\
\nonumber
G_{ECot} \sim \frac{h}{8e^2} \sum_{\sigma} 
\,G_{1\sigma}G_{2\sigma}\;\frac{e^{-2 r/ \pi \xi_{\omega}}}{(k_Fr)^2}
\end{eqnarray}

Here $G_{1\sigma}$ and $G_{2-\sigma}$ are the one-electron conductances in 
the normal state for a given spin. As shown in 
refs. \cite{DF,falci,melin1}, 
both crossed conductances can be distinguished and measured as soon as 
leads 1 and 2 are spin-polarized. A recently proposed alternative is to use 
the crossed correlations of shot noise \cite{pistolesi}.

The dimensionality of the superconductor is crucial : the BCS coherence 
length for a clean superconductor can be quite large, and the algebraic 
factor
describing the ballistic propagation of quasiparticles in S is the most 
limiting effect in three dimensions. The constraint is weaker in two 
dimensions
where it becomes 
$\sim \frac{e^{-2 r/ \pi \xi_{\omega}}}{k_Fr}$, and in one dimension where 
one finds $e^{-2 r/ \pi \xi_{\omega}}$ \cite{recher}. 

\section{Case of a diffusive superconductor}
Disorder is always present in low dimensional superconductors (films) thus 
it is important to consider the diffusive limit where elastic 
scattering occurs with a mean-free path $l$. We assume no spin scattering 
that could be due to magnetic impurities or spin-orbit interaction.
Using the golden rule or Keldysh technique, one can generalize Eqs. 
(2-5) for any realization 
of the disorder by replacing the bare Green's functions $g$ and $f$ by the 
ones dressed by impurity scattering. On the other hand, disorder 
averaging implies to perform the average on the products of retarded and advanced 
Green's functions $g^rg^a$ and 
$f^rf^a$. These averages are related to 
the normal 
and the anomalous integrated diffusion probabilities \cite{livre}, ${\cal P} (r) = 
\int_{-\infty}^{\infty}\;dt\;{\cal P} (\vec{r}_{1},\vec{r}_{2},t)$ for $r=r_{12}$.

\begin{eqnarray}
{\cal P} (r_{12})\;=\;\frac{1}{2\pi\rho_0} \;\; 
\overline{g^r_{\sigma}(r_{12},\omega)\;g^a_{\sigma}(r_{21},\omega)}\\
\tilde{{\cal P}} (r_{12})\;=\;\frac{1}{2\pi\rho_0} \;\; 
\overline{f^r_{\sigma}(r_{12},\omega)\;f^a_{\sigma}(r_{21},\omega)}
\end{eqnarray}

\noindent
taken at $\omega \sim \varepsilon_F$. $\rho_0$ is the normal state density of states in the 
superconductor, and ${\cal P} (r)$ corresponds to
 the diffuson in a normal metal (electron-electron channel), and 
$\tilde{{\cal P}} (r)$ 
is the analogue with normal propagators replaced by anomalous ones (see Figure 2). 
The former describes the virtual diffusion (below the gap) of an out-of-
equilibrium quasiparticle, electron or hole. The second describes the 
anomalous diffusion of an electron becoming a hole (with emission of a 
Cooper pair) or vice-versa. 

\begin{figure}
\includegraphics[scale=.5]{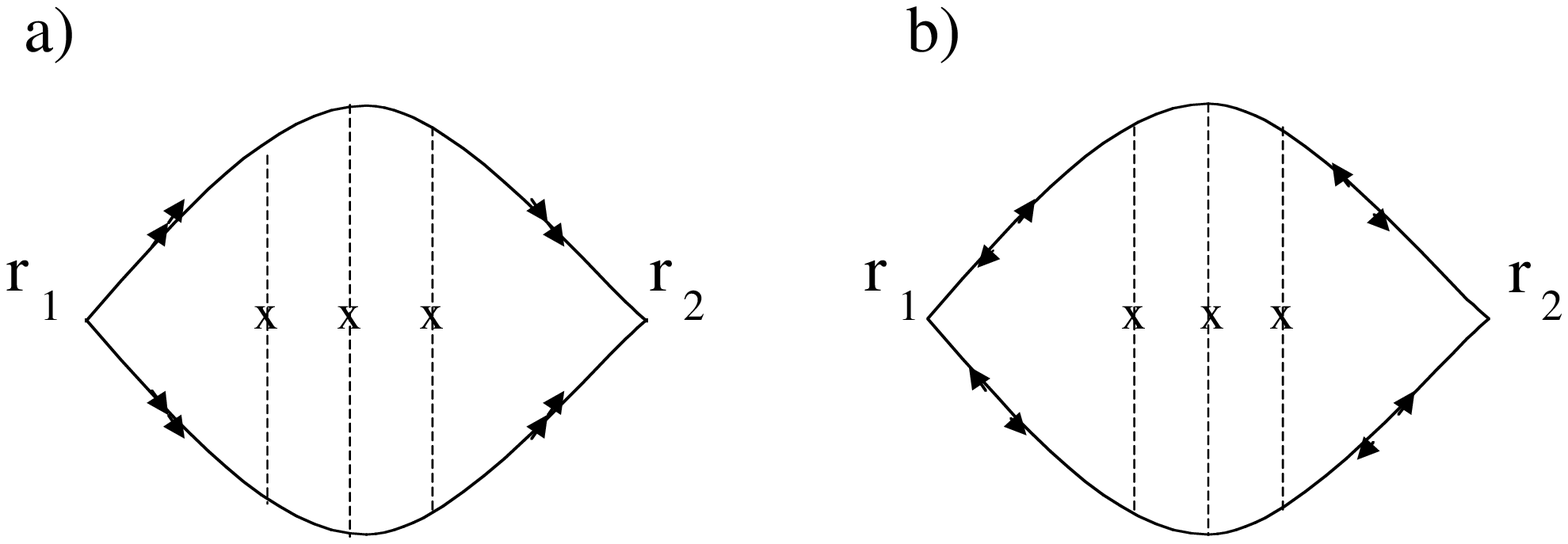}
\caption{The diffuson diagrams in the superconductor, showing multiple impurity scattering 
of quasiparticles in the normal (a) and the anomalous (b) channel. Continuous lines denote the 
propagators $g$ (a)
and $f$ (b), dotted lines the impurity vertices.} 
\end{figure}

The solution for the diffusons starts from the Drude-Boltzmann 
approximation, where the $g$'s and $f$'s are independently averaged on 
disorder

\begin{eqnarray}
{\cal P}_0 (r_{12})\;=\;\frac{1}{2\pi\rho_0} \;\; 
\overline{g^r_{\sigma}(\omega,r_{12})}\;\;\;\overline{g^a_{\sigma}(\omega,r_{21})}\\
\tilde{{\cal P}}_0 (r_{12})\;=\;\frac{1}{2\pi\rho_0} \;\; 
\overline{f^r_{\sigma}(\omega,r_{12})}\;\;\;\overline{f^a_{\sigma}(\omega,r_{21})}
\end{eqnarray}

\noindent
where $\overline{g^{r,a}(\omega,r)}$ and 
$\overline{f^{r,a}(\omega,r)}$ are obtained from the propagators 
$g^{r,a}(\omega,r)$ and $f^{r,a}(\omega,r)$ in the clean superconductor 
case, e.g. $\overline{g^{r,a}(\omega,r)} = 
g^{r,a}(\omega,r) e^{-r/2l}$, 
$\overline{f^{r,a}(\omega,r)} = f^{r,a}(\omega,r) e^{-r/2l}$ 
\cite{Abrikosov}. 

The full diffusons are obtained from the integral equation \cite{livre}

\begin{eqnarray}
{\cal P} (\vec{r}_1,\vec{r}_2))\;=\;2\pi\rho_0\;\;\int\;d\vec{r}'d\vec{r}''\;\;{\cal P}_0 
(\vec{r}_1,\vec{r}')\;\Gamma(\vec{r}',\vec{r}'')\;{\cal P}_0 (\vec{r}'',\vec{r}_2)\\
\tilde{{\cal P}} (\vec{r}_1,\vec{r}_2)\;=\;2\pi\rho_0\;\;\int\;d\vec{r}'d\vec{r}''\;\;\tilde{{\cal 
P}}_0 (\vec{r}_1,\vec{r}')\;\tilde{\Gamma}(\vec{r}',\vec{r}'')\;\tilde{{\cal P}}_0 (\vec{r}'',\vec{r}_2)
\end{eqnarray}

\noindent
the vertex function $\Gamma$ obeying

\begin{eqnarray}
\Gamma(\vec{r}_1,\vec{r}_2)\;=\;\gamma_e \delta(\vec{r}_1-\vec{r}_2) + 
\frac{1}{\tau_e}\;\;\int\;d\vec{r}'\;\;\Gamma(\vec{r}_1,\vec{r}')\;{\cal P}_0 (\vec{r}',\vec{r}_2)\\
\tilde{\Gamma}(\vec{r}_1,\vec{r}_2)\;=\;\gamma_e \delta(\vec{r}_1-\vec{r}_2) + 
\frac{1}{\tau_e}\;\;\int\;d\vec{r}'\;\;\tilde{\Gamma}(\vec{r}_1,\vec{r}')\;\tilde{{\cal P}}_0 
(\vec{r}',\vec{r}_2)
\end{eqnarray}

where $\gamma_e = (2\pi\rho_0\tau_e)^{-1}$ is the bare vertex and and 
$\tau_e^{-1} = 2\pi \rho_0 n_i |v_i|^2 = \frac{v_F}{l}$ the inverse 
scattering time 
for an density $n_i$ of impurities with potential strength $v_i$. 

Let us consider the dirty limit $l < \xi_0$, which means that the 
quasiparticle encounter many collisions before decaying. 
Then ${\cal P}_0$, $\tilde{{\cal P}}_0$ decay on $l$ while $\Gamma$ a 
priori decays more slowly, allowing a gradient 
approximation

\begin{eqnarray}
\Gamma(\vec{r}_1,\vec{r}_2)\;\sim\;\gamma_e \delta(\vec{r}_1-\vec{r}_2) + 
\frac{1}{\tau_e}\;\Gamma(\vec{r}_1,\vec{r}_2)\;\langle{\cal P}_0 (r)\rangle\;
+\;\nabla^2\Gamma(\vec{r}_1,\vec{r}_2)\;\langle r^2{\cal P}_0 (r)\rangle\\
\tilde{\Gamma}(\vec{r}_1,\vec{r}_2)\;\sim\;\gamma_e \delta(\vec{r}_1-\vec{r}_2) + 
\frac{1}{\tau_e}\;\tilde{\Gamma}(\vec{r}_1,\vec{r}_2)\;\langle{\tilde{\cal P}}_0 
(r)\rangle\;
+\;\nabla^2\tilde{\Gamma}(r_1,r_2)\;\langle r^2\tilde{{\cal P}}_0 
(r)\rangle
\end{eqnarray}

\noindent
One easily calculates 

\begin{eqnarray}
\langle{\cal P}_0 (r)\rangle\;=\;\langle{\tilde{\cal P}}_0 
(r)\rangle\;=\;\tau_e\frac{\Delta^2}{\Delta^2-\omega^2}\;\frac{1}{1+\frac{l}{\xi_{\omega}}}\\
\langle r^2{\cal P}_0 (r)\rangle\;=\;\langle r^2\tilde{{\cal P}}_0 
(r)\rangle\;=
\;2\tau_e\frac{\Delta^2}{\Delta^2-\omega^2}\;\frac{l^2}{(1+\frac{l}{\xi_{\omega}})^3}
\end{eqnarray}

\noindent
This leads to the solution, valid for $r >> l$

\begin{equation}
{\cal P}(r) = \tilde{{\cal P}}(r) = 
(1+\frac{l}{\xi_{\omega}})\;\;\frac{\Delta^2}{\Delta^2-\omega^2}\;\;
\frac{1}{4\pi D r}e^{-r/\tilde{\xi}_{\omega}}
\end{equation}

\noindent
with 

\begin{equation}
\tilde{\xi}_{\omega}^{-2} = 
(D\tau_e)^{-1}\;(1+\frac{l}{\xi_{\omega}})^2\;[\frac{l}{\xi_{\omega}} - 
\frac{\omega^2}{\Delta^2} - 
\frac{l}{\xi_{\omega}}\frac{\omega^2}{\Delta^2}]
\end{equation}

\noindent
To lowest order in $\frac{l}{\xi_{\omega}} $ and $\frac{\omega}{\Delta}$,  
one finds 

\begin{equation}
\tilde{\xi}_{\omega} \sim \sqrt{\frac{D\tau_e \xi_{\omega}}{l}} \sim 
\sqrt{l \xi_{\omega}}
\end{equation}

\noindent
justifying a posteriori the above gradient approximation. 

We thus find that in the dirty limit, the range of the 
diffusons, thus of the non-local scattering probabilities, 
is reduced only to the "dirty limit" coherence length, and not to the 
mean-free-path. As for $\xi_{\omega}$, it diverges as $\omega$ approaches 
the superconducting gap. 

We can now write the crossed conductances

\begin{eqnarray}
G_{CAnd} \sim \frac{h}{8e^2} \sum_{\sigma} 
\,G_{1\sigma}G_{2-\sigma}\;\frac{e^{-r/ \tilde{\xi}_{\omega}}}{\hbar 
\rho_0 Dr}\\
\nonumber
G_{ECot} \sim \frac{h}{8e^2} \sum_{\sigma} 
\,G_{1\sigma}G_{2\sigma}\;\frac{e^{-r/ \tilde{\xi}_{\omega}}}{\hbar \rho_0 
Dr}
\end{eqnarray}

Besides the smaller decay length, one notices the different algebraic 
dependence, in $1/r$ instead of $1/r^2$ for the clean limit. In more 
detail, 
the conductances vary like $\frac{1}{(k_Fr)(k_Fl)}e^{-r/\tilde{\xi}}$, 
showing that for $l < r < \tilde{\xi}$, the dirty case is more favourable 
than the clean one. This result holds when all the dimensions of the 
superconductor are larger than $l$. If one of them is smaller (very thin 
film), diffusion 
becomes two-dimensional and the solution of the diffusion equation leads 
to a dependence $\frac{1}{\sqrt{r}}e^{-r/\tilde{\xi}}$ if 
$r > \tilde{\xi}$ and $-ln(\frac{r}{\tilde{\xi}})$ if $r < \tilde{\xi}$, 
again showing the advantage of diffusive behaviour. 

One can use this result to evaluate the conductance for extended contacts 
$1$ and $2$. From Eqs. (2-5) it is given approximately by

\begin{equation}
G_{CAnd,ECot} \sim \int\;d\vec{r}_1\;d\vec{r}_2\;\;G_{CAnd,ECot}(r_{12})
\end{equation}

\noindent
As an example, for two linear contacts facing each other at a distance $R 
< \tilde{\xi}$, of length and width much larger than $\tilde{\xi}$ (Figure 1b), 
one easily finds that the $1/r$ factor integrates out and $G_{CAnd,ECot} \sim 
e^{-R/\tilde{\xi}}$. 

To summarize, we have shown that Andreev and cotunneling processes between 
distinct tunneling contacts on a dirty superconductor decay on the 
coherence length $\tilde{\xi} = \sqrt{l\xi_0}$, and that the algebraic 
prefactor decreases like $1/r$ with the contact distance instead of
$1/r^2$ in the clean case. For extended contacts closer than $\xi$ the 
crossed conductances can be more easily observed.

\bigskip
The author is grateful to G. Montambaux and G. 
Deutscher for stimulating discussions. After completing this work the 
author was informed about an equivalent calculation of 
crossed Andreev amplitude with a dirty superconductor 
\cite{chtchelkatchev}. LEPES is under convention with Universit\'e Joseph 
Fourier.

\end{document}